# POSSIBILITY OF HOLE CONFINEMENT IN THE WEAK COUPLING REGIME OF ONE DIMENSIONAL ISING ANTIFERROMAGNET AND THE EMERGENCE OF SPIN-CHARGE SEPARATION IN THE BULK LIMIT.

S. Ehika [a*], O.R. Okanigbuan[a] and J.O.A. Idiodi[b]

[a]Department of Physics, Ambrose Alli University Ekpoma, Edo State, Nigeria.

[b]Department of Physics, University of Benin, Benin City, Nigeria.

*Corresponding author.

E-mail address: ehikasimon@yahoo.com

Postal address: Department of Physics, Ambrose Alli University, P.M.B 14, Ekpoma, Edo State, Nigeria

Telephone: +2348074478911

## ABSTRACT

The energy of one hole in one dimensional Ising antiferromagnet is calculated using exact diagonalization (ED) method on finite systems with number of sites N ranging from 4 to 50. This study gives a power-law description for the hole energy in the weak coupling region ($0.001 \leq J_z/t \leq 0.1$). This power-law description can otherwise be understood as a weak "string-like" energy that tends to compromise the motion of the hole by confining it to its "birth" site. However, for large N, the hole is found to escape from this confinement and hence propagates as a free particle in agreement with theoretical results in the bulk limit and experimental observations from angle resolved photoemission spectroscopy (ARPES) of spin-charge separation. In the strong coupling regime ($J_z/t \gg 1$), the energy of the hole is found to be independent of the system size. Accordingly, the velocities corresponding to charge and spin degrees of freedom in this regime become comparable. Under this situation, the hole may be treated as a localized particle.

## 1. Introduction

The behaviour of the cuprates unlike the conventional superconductors is known to be governed by strong electronic interactions. This has spurred interest in the field of strongly correlated electron systems with the hope that it will unravel the mystery behind superconductivity in the cuprates and possibly solve the problems inherent in these materials. The cuprates are also classified as Mott insulators with dominant physical processes (charge transport, antiferromagnetic exchange, etc) that participate in the formation of the superconducting condensate located in the copper-oxygen planes [1-3].

At half filling, hopping of electrons is highly forbidden due to the strong onsite electronic repulsion. The magnetic properties of these half-filled (undoped) systems are well described by the isotropic spin-1/2 Heisenberg model [4]. Numerical and analytical studies have shown that the ground state properties of Heisenberg systems are sensitive to the geometry of the lattice and the distribution of the exchange couplings between two neighbouring spins [5, 6].

It has been observed that under light doping, which removes electrons thereby producing mobile holes in the $CuO_2$ planes, antiferromagnetic ordering is destroyed and the compound becomes superconducting [7]. There have been rigorous and intense researches on one and quasi-one dimensional (1D and quasi -1D) systems. The motivation for researches on 1D and

1D and quasi -1D systems is because of the intriguing phenomena that have been discovered theoretically and experimentally in them. It will be worthwhile to review some of these phenomena. For instance, the injection of a single hole into quasi-one dimensional Mott insulators such as $Sr_2CuO_3$ and $SrCuO_2$ gives rise to Spin-charge separations [8-11]. This discovery therefore gives birth to two quasiparticles namely, holon (possessing only charge) and spinon (possessing only spin). More so, an experiment with $Sr_2CuO_3$ found an orbiton liberating itself from spinons and propagating through the lattice as a distinct quasi-particle with a dispersion of ~ 0.2eV [12]. This observation of a free orbiton is another example of particle fractionalization known as Spin-orbital separation. In ref. [13], further fractionalization of spin and charge beyond the Luttinger-liquid paradigm was discovered. An evidence for quasi-particle break down in the quasi-one dimensional Ising ferromagnet has been presented theoretically and experimentally in $CoNb_2O_6$ [14]. A transition from gapless edge states in a noninteracting topological band insulator to spinon edge states in a topological Mott insulator has been observed theoretically [15]. These reviews point out the relevance of 1D and quasi-1D cuprates in the scheme of high-Tc superconductors.

Away from 1D systems, low dimensional systems such as ladder systems have been used as a fantastic playground to observe the effect of hole dynamics on antiferromagnetically ordered spin background. For instance, self-localization that invalidates the quasi-particle picture has been observed in antiferromagnetic ladder systems [16]. This observation of self-localization of a single hole also validates an earlier report of spinon confinement in a coupled chain [17]. This self-localization of a single hole which arises from quantum destructive interference of the phase string signs hidden in the $t-J$ ladders can be removed by pairing two holes [18]. A study of hole motion in a classical Neél state in two dimensions in Refs. [19-21], concluded that the motion of the hole is confined. By considering quantum spin fluctuations and some complicated paths, a single hole was made mobile [22-25]. But recent discovery of string excitations in refs.[26-28] from ARPES studies of cuprates seems to be in support of the string picture proposed in [21] in the intermediate coupling. It is therefore obvious that there is lack of consensus regarding the dynamics of a hole in an antiferromagnet. This field is still evolving with theoretical and experimental researches geared towards addressing some of the problems of this hole dynamics. The aim of the present work is to study the propagation of a single hole in 1D $t-J_z$ (Ising) model with exact diagonalization method. It is hoped that this simplified model will shed more light on some elusive aspect of hole dynamics and provide further evidence of spinon-holon decoupling in the thermodynamic limit. The rest of the paper is organized as follows: In section 2, the Ising model in 1D is presented and an effective Hamiltonian describing the hole dynamics in the midst of the Ising interaction term is introduced. In section 3, we present exact diagonalization of some finite systems. In section 4, we present and discuss the numerical ground state energy of the hole as a function of $J_z/t$. Observation of weak coupling between the holon and spinon for finite systems and its subsequent decoupling in the bulk limit is also presented in this section. We present experimental realization of spin-charge separation in 1D cuprates in section 5. A brief conclusion will close the paper in section 6.

## 2. Hole motion in one dimensional Ising antiferromagnet

The Ising ($t-J_z$) model is the strongly anisotropic limit of the $t-J$ model which captures some general properties of the doped antiferromagnets (AF). The absence of spin fluctuations in the Ising model simplifies the analytical treatment of this problem and makes it possible to visualize the independent effect of hole(s) on the antiferromagnet. It has been established that the undoped cuprates display conventional Neel order which can be described by quantum Heisenberg model. When doped, the long-range antiferromagnetic order is destroyed and superconductivity appears. This disappearance of antiferromagnetic ordering shows that the carries (holes) are not static but mobile. The central issue is whether a single hole can propagate freely (coherently) in an antiferromagnetic background. Although, there are finite concentration of charge carriers in real superconductors, but this work will limit itself on the dynamics of a single hole in one dimensional Ising antiferromagnet. It is hoped that the solution to this problem will shed more light on the underlying Physics of high temperature superconductors (HTS). The Hamiltonian describing hole hopping in an Ising model is a simple generalization of the $t-J$ model where only the $S^z$ component of the spins are retained. In one dimension this model reads

$$H = -t\sum_{i\sigma}\left[C^\dagger_{i\sigma}C_{i+1\sigma} + C^\dagger_{i+1\sigma}C_{i\sigma}\right] + J_z\sum_{i} S^z_i S^z_{i+1} . \tag{1}$$

Here, $C^\dagger_{i\sigma}$ ($C_{i\sigma}$) is the creation (annihilation) operator of an electron with spin σ at the lattice site $i$; $S^z_i = 1/2(c^\dagger_{i\uparrow}c_{i\uparrow} - c^\dagger_{i\downarrow}c_{i\downarrow})$ is the Ising component of the antiferromagnetic spin interaction; $t$ is the effective transfer integral (or the kinetic energy term) and $J_z$ is the antiferromagnetic exchange energy for a pair of nearest neighbour spins. In this model, the constraint of no double occupancy is understood.

In this model, even chains subject to periodic boundary conditions are considered. At half filling, the ground state is a classical Neél state $|N\rangle$. Out of the two states obtained by translation of one lattice vector, the one with a spin down at the origin is selected. The annihilation of the spin down electron at the origin defines the starting state, i.e. $|0\rangle = c_{0\downarrow}|N\rangle$. The action of the Hamiltonian in eqn. (1) on $|0\rangle$ will generate states which can uniquely be labeled by the position of the hole $|R\rangle$. In the Bulk limit, close paths of the hole along the ring can be neglected [19]. On account of this, Sorella and Parola in Ref. [29] proposed the following Hamiltonian for the dynamics of a hole in the Ising limit of 1D cuprates:

$$H|R\rangle = -t[|R-1\rangle + |R+1\rangle] + \left(E_N + \frac{J}{2}\right)|R\rangle - \frac{J}{2}\delta_{R,0}|R\rangle,\tag{2}$$

where $E_N$ is the energy of the Neél state. The diagonalized Hamiltonian matrix in this subspace can be approximated by

$$E_b = E_N + \frac{1}{2}\left(J - \sqrt{J^2 + 16t^2}\right).\tag{3}$$

In this work, the following Hamiltonian for a hole in finite 1D cuprates is proposed.

$$H'|R\rangle = -t[|R-1\rangle + |R+1\rangle] + (J_z)|R\rangle - \frac{J_z}{2}\delta_{R,0}|R\rangle,\tag{4}$$

where

$$H' = H - H_{ising}.\tag{5}$$

Here, $H_{ising}$ is the Ising Hamiltonian of the Neél state. The emergence of (4) arises from the fact that the ferromagnetic bond or spinon created as the hole propagates in finite 1D antiferromagnet is not static, but mobile. In (2), the ferromagnetic bond is approximated to be static in the thermodynamic limit. The energy of a single hole $E_N$ relative to the Neél background is given by

$$E_h = \langle R|H|R\rangle - E_N.\tag{6}$$

The translational invariance property of the $t - J_z$ model will be utilized in reducing the size of the Hilbert space of a single hole. This means that any one hole state with definite momentum k and spin ↑ can be written as

$$|\psi_k\rangle = \frac{1}{\sqrt{L}}\sum_{R=0}^{L-1} e^{-ikR} T_L^R c_{0\downarrow}|\sigma_0\rangle,\tag{7}$$

where $|\sigma_0\rangle$ is a suitable spin state with the spin at the origin $R=0$ fixed to ↓ and $T_L$ is the spin translation operator defined by the transformation property

$$T_L S_R T_L^{-1} = S_{R+1},\tag{8}$$

where periodic boundary conditions (PBC) over the L sites are assumed in order to define the effect of translation at the rightmost site. With translational symmetry the Hilbert space of size N(N-1) is reduced by a factor of N according to the equation

$$S_N = \frac{N(N-1)}{N} = N-1. \tag{9}$$

## 3. Exact diagonalization of finite chains

In this section, exact diagonalization of some finite systems with PBC will be carried.

### 3.1. Four sites chain

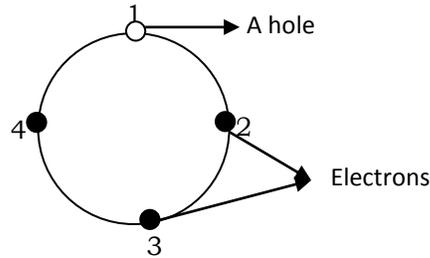

Fig.1. A four sites chain with a hole subject to periodic boundary conditions. The topology of the spin space is that of a circle.

The Neél states of this system obtained by translation of one lattice vector are $|1\uparrow,2\downarrow 3\uparrow,4\downarrow\rangle$ and $|1\downarrow 2\uparrow,3\downarrow,4\uparrow\rangle$.

Since we are interested in the subspace of $S^z_{tot} = 1/2$, the annihilation of say the down spin electron at site 1 defines our starting state. This state denoted by $|0\rangle$ is given by $|0\rangle = |1,2\uparrow,3\downarrow,4\uparrow\rangle$. By implementing the translational symmetry discussed in section 2, the Hilbert space of this system of size 12 is reduced to 4 according to (9) as shown below

$$|\phi_0\rangle = \frac{1}{2}\left[|1,2\uparrow,3\downarrow,4\uparrow\rangle + |1\uparrow,2,3\uparrow,4\downarrow\rangle + |1\downarrow,2\uparrow,3,4\uparrow\rangle + |1\uparrow,2\downarrow,3\uparrow,4\rangle\right]$$

$$|\phi_1\rangle = \frac{1}{2}\left[|1,2\downarrow,3\uparrow,4\uparrow\rangle + |1\uparrow,2,3\downarrow,4\uparrow\rangle + |1\uparrow,2\uparrow,3,4\downarrow\rangle + |1\downarrow,2\uparrow,3\uparrow,4\rangle\right]$$

$$|\phi_2\rangle = \frac{1}{2}\left[|1,2\uparrow,3\uparrow,4\downarrow\rangle + |1\downarrow,2,3\uparrow,4\uparrow\rangle + |1\uparrow,2\downarrow,3,4\uparrow\rangle + |1\uparrow,2\uparrow,3\downarrow,4\rangle\right]$$

Application of $H'$ to these bases states gives

$$H'|\phi_1\rangle = -t|\phi_0\rangle - t|\phi_2\rangle + J_z|\phi_1\rangle$$

$$H'|\phi_0\rangle = -t|\phi_1\rangle - t|\phi_2\rangle + \frac{J_z}{2}|\phi_0\rangle$$

$$H'|\phi_2\rangle = -t|\phi_0\rangle - t|\phi_1\rangle + J_z|\phi_2\rangle$$

The Hamiltonian matrix becomes

$$H' = \begin{bmatrix} J_z/2 & -t & -t \\ -t & J_z & -t \\ -t & -t & J_z \end{bmatrix}. \qquad (10)$$

The complete diagonalization of this matrix will give the ground state energy

$$E_g = \frac{1}{4}\left[3J_z - 2t - \sqrt{J_z^2 - 4J_z t + 36t^2}\right]. \qquad (11)$$

Following the procedure outline for generating bases vectors as demonstrated for four-site chain and by utilizing sthe translational invariance of the $t-J_z$, the resulting Hamiltonian matrices for 6-, 8-, 10- and 20-site chains are:

**3.2. Six –site chain**

$$H' = \begin{bmatrix} \frac{J_z}{2} & -t & -t & 0 & 0 \\ -t & J_z & 0 & -t & 0 \\ -t & 0 & J_z & 0 & -t \\ 0 & -t & 0 & J_z & -t \\ 0 & 0 & -t & -t & J_z \end{bmatrix}. \qquad (12)$$

### 3.3. Eight-site chain

$$H' = \begin{bmatrix} \frac{J_z}{2} & -t & -t & 0 & 0 & 0 & 0 \\ -t & J_z & 0 & -t & 0 & 0 & 0 \\ -t & 0 & J_z & 0 & -t & 0 & 0 \\ 0 & -t & 0 & J_z & 0 & -t & 0 \\ 0 & 0 & -t & 0 & J_z & 0 & -t \\ 0 & 0 & 0 & -t & 0 & J_z & -t \\ 0 & 0 & 0 & 0 & -t & -t & J_z \end{bmatrix}. \tag{13}$$

### 3.4. Ten-site chain

$$H' = \begin{bmatrix} \frac{J_z}{2} & -t & -t & 0 & 0 & 0 & 0 & 0 & 0 \\ -t & J_z & 0 & -t & 0 & 0 & 0 & 0 & 0 \\ -t & -t & J_z & 0 & -t & 0 & 0 & 0 & 0 \\ 0 & 0 & 0 & J_z & 0 & -t & 0 & 0 & 0 \\ 0 & 0 & -t & 0 & J_z & 0 & -t & 0 & 0 \\ 0 & 0 & 0 & -t & 0 & J_z & 0 & -t & 0 \\ 0 & 0 & 0 & 0 & -t & 0 & J_z & 0 & -t \\ 0 & 0 & 0 & 0 & 0 & -t & 0 & J_z & -t \\ 0 & 0 & 0 & 0 & 0 & 0 & -t & -t & J_z \end{bmatrix}. \tag{14}$$

### 3.5. Twenty-site chain

$$H' = \begin{bmatrix}
\frac{J_z}{2} & -t & -t & 0 & 0 & 0 & 0 & 0 & 0 & 0 & 0 & 0 & 0 & 0 & 0 & 0 & 0 & 0 & 0 \\
-t & J_z & 0 & -t & 0 & 0 & 0 & 0 & 0 & 0 & 0 & 0 & 0 & 0 & 0 & 0 & 0 & 0 & 0 \\
-t & -t & J_z & 0 & -t & 0 & 0 & 0 & 0 & 0 & 0 & 0 & 0 & 0 & 0 & 0 & 0 & 0 & 0 \\
0 & 0 & 0 & J_z & 0 & -t & 0 & 0 & 0 & 0 & 0 & 0 & 0 & 0 & 0 & 0 & 0 & 0 & 0 \\
0 & 0 & -t & 0 & J_z & 0 & -t & 0 & 0 & 0 & 0 & 0 & 0 & 0 & 0 & 0 & 0 & 0 & 0 \\
0 & 0 & 0 & -t & 0 & J_z & 0 & -t & 0 & 0 & 0 & 0 & 0 & 0 & 0 & 0 & 0 & 0 & 0 \\
0 & 0 & 0 & 0 & -t & 0 & J_z & 0 & -t & 0 & 0 & 0 & 0 & 0 & 0 & 0 & 0 & 0 & 0 \\
0 & 0 & 0 & 0 & 0 & -t & 0 & J_z & 0 & -t & 0 & 0 & 0 & 0 & 0 & 0 & 0 & 0 & 0 \\
0 & 0 & 0 & 0 & 0 & 0 & -t & 0 & J_z & 0 & -t & 0 & 0 & 0 & 0 & 0 & 0 & 0 & 0 \\
0 & 0 & 0 & 0 & 0 & 0 & 0 & -t & 0 & J_z & 0 & -t & 0 & 0 & 0 & 0 & 0 & 0 & 0 \\
0 & 0 & 0 & 0 & 0 & 0 & 0 & 0 & -t & 0 & J_z & 0 & -t & 0 & 0 & 0 & 0 & 0 & 0 \\
0 & 0 & 0 & 0 & 0 & 0 & 0 & 0 & 0 & -t & 0 & J_z & 0 & -t & 0 & 0 & 0 & 0 & 0 \\
0 & 0 & 0 & 0 & 0 & 0 & 0 & 0 & 0 & 0 & -t & 0 & J_z & 0 & -t & 0 & 0 & 0 & 0 \\
0 & 0 & 0 & 0 & 0 & 0 & 0 & 0 & 0 & 0 & 0 & -t & 0 & J_z & 0 & -t & 0 & 0 & 0 \\
0 & 0 & 0 & 0 & 0 & 0 & 0 & 0 & 0 & 0 & 0 & 0 & -t & 0 & J_z & 0 & -t & 0 & 0 \\
0 & 0 & 0 & 0 & 0 & 0 & 0 & 0 & 0 & 0 & 0 & 0 & 0 & -t & 0 & J_z & 0 & -t & 0 \\
0 & 0 & 0 & 0 & 0 & 0 & 0 & 0 & 0 & 0 & 0 & 0 & 0 & 0 & -t & 0 & J_z & 0 & -t \\
0 & 0 & 0 & 0 & 0 & 0 & 0 & 0 & 0 & 0 & 0 & 0 & 0 & 0 & 0 & -t & 0 & J_z & -t \\
0 & 0 & 0 & 0 & 0 & 0 & 0 & 0 & 0 & 0 & 0 & 0 & 0 & 0 & 0 & 0 & -t & -t & J_z
\end{bmatrix} \quad (15)$$

The results for the numerical ground state energy corresponding to these matrices for different values of the coupling strength $J_z/t$ will be presented in section 4.

### 3.6. Fifty-site chain

In similar way, the matrix corresponding to fifty sites chain can be generated. The matrix arising from this system is large (i.e 49x49 with symmetry consideration). Hence, only the results for the numerical ground state energy will be presented in section 4.

### 4. Results and discussion
### 4.1. Numerical result for exact diagonalization of finite chains

This section presents the numerical exact result for the energy of a single hole for 4-, 6-, 8-, 10-, 20- and 50 site chains. From Table I, it is observed that at $J_z/t = 0$ the energy of the hole is -2. At this energy, the hole can propagate freely through the antiferromagnetic background without disrupting the spin background and expending energy. This is because the antiferromagnetic coupling parameter $J_z$ which provides the confining potential is zero. The slight increase in the energy of the hole as N is increased is due to the formation of "string" of paired ferromagnetically aligned up spins (spinon) coupled weakly to the hole. This energy cost will result to a linear rising potential that tends to confine the hole to its "birth site". Hence, the coherent propagation of the hole is slightly compromised. Also, as observed from Table I, the hole energy tends towards a constant value as N is increased from 4 to 50. This behaviour at large N is the signature of spin-charge separation as observed theoretically in the thermodynamic limit in Ref. [29] and experimentally in Ref. [8]. Also, the energy of the hole is found to increase almost linearly with $J_z/t$ because of the increase in magnetic energy expended by the hole. This approximate linear behaviour of the energy expended by the hole as the magnetic coupling constant is increased is captured by Fig. 2.

Table I. The energy $E_h/t$ of a hole as a function of $J_z/t$ for 4, 6, 8, 10, 20 and 50 sites. The table shows an increase in the energy of the hole as N is increased.

| $J_z/t$ | $E_h/t$ (4 sites) | $E_h/t$ (6 sites) | $E_h/t$ (8 sites) | $E_h/t$ (10 sites) | $E_h/t$ (20 sites) | $E_h/t$ (50 sites) |
|---|---|---|---|---|---|---|
| 0.00000 | -2.00000 | -2.00000 | -2.00000 | -2.00000 | -2.00000 | -2.00000 |
| 0.01000 | -1.99167 | -1.99100 | -1.99072 | -1.99056 | -1.99027 | -1.99010 |
| 0.02000 | -1.98334 | -1.98201 | -1.98144 | -1.99112 | -1.98053 | -1.98021 |
| 0.05000 | -1.95838 | -1.95505 | -1.95362 | -1.95283 | -1.95137 | -1.95057 |
| 0.10000 | -1.91685 | -1.91020 | -1.90735 | -1.90577 | -1.90285 | -1.90127 |
| 0.20000 | -1.83408 | -1.82082 | -1.81514 | -1.81198 | -1.80621 | -1.80321 |
| 0.40000 | -1.66969 | -1.64338 | -1.63213 | -1.62592 | -1.61483 | -1.61026 |
| 0.60000 | -1.50688 | -1.46782 | -1.45121 | -1.44214 | -1.42675 | -1.42243 |
| 0.80000 | -1.34568 | -1.29428 | -1.27264 | -1.26101 | -1.24283 | -1.23962 |
| 1.00000 | -1.18614 | -1.12291 | -1.09665 | -1.08290 | -1.06362 | -1.06155 |
| 1.50000 | -0.79473 | -0.70479 | -0.66958 | -0.66958 | -0.63649 | -0.63600 |
| 2.00000 | -0.41421 | -0.30278 | -0.26308 | -0.24698 | -0.23616 | -0.23607 |
| 2.50000 | -0.04473 | 0.08215 | 0.12192 | 0.13520 | 0.14149 | 0.14151 |
| 3.00000 | 0.31386 | 0.45002 | 0.48662 | 0.49655 | 0.49999 | 0.50000 |
| 4.00000 | 1.00000 | 1.13919 | 1.16576 | 1.17056 | 1.17157 | 1.17157 |

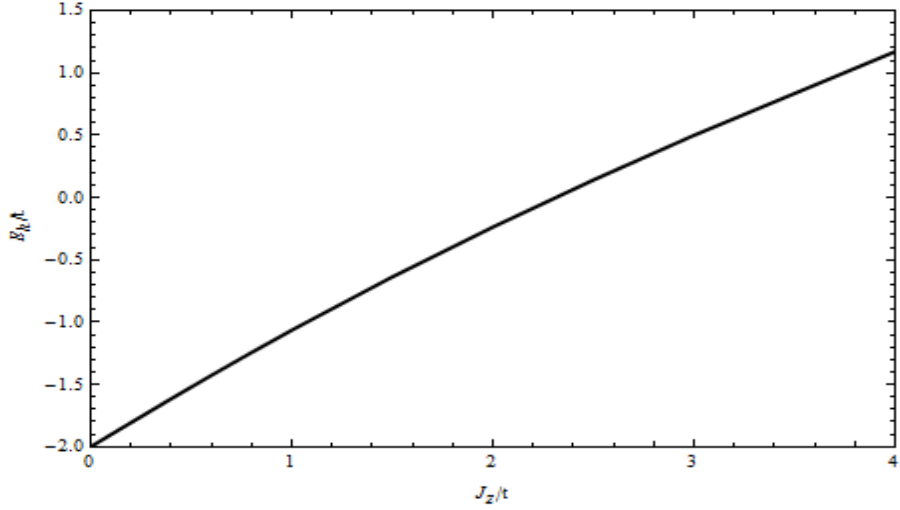

Fig.2. Hole energy $E_h/t$ as function of $J_z/t$ for one dimensional Ising antiferromagnet. From the figure, an increase in $J_z$ increases the magnetic energy expended by the hole.

### 4.2. Observation of string energy in the weak coupling regime

In this section, the possibility of string excitations in the weak coupling regime of finite chains is discussed. The behaviour of these string energies in the bulk limit is also discussed. In Fig.3, the energy of the hole in the weak coupling regime ($0.01 \leq J_z/t \leq 0.1$) is described by a power law similar to the string energy proposed in [21]. This power law behaviour for 4-, 6-, 8-, 10-, 20- and 50-site chains are respectively shown in eqns. (16), (17), (18) (19) (20) and (21)

$$E_h/t = -1.999 + 0.77(J_z/t)^{0.97} \tag{16}$$

$$E_h/t = -1.999 + 0.77(J_z/t)^{0.94} \tag{17}$$

$$E_h/t = -1.999 + 0.77(J_z/t)^{0.93} \tag{18}$$

$$E_h/t = -1.999 + 0.77(J_z/t)^{0.92} \tag{19}$$

$$E_h/t = -1.999 + 0.77(J_z/t)^{0.9} \tag{20}$$

$$E_h/t = -1.999 + 0.77(J_z/t)^{0.892} \tag{21}$$

For small distances (small N), the energy of the hole shows a linear dependence on the length of the string due to increase in the string potential. At large distances or large N (corresponding to the thermodynamic or bulk limit), this dependence breaks down as observed from the exponent of ($J_z/t$) tending towards a constant value. This means that the coupling between the hole and the ferromagnetic bond representing the spinon is broken, thereby giving the hole a coherent propagation. This behaviour which is an evidence for spin-charge separation has been realized experimentally in Refs. [8-10]. The dependence of $E_h/t$ on N for fixed values of the coupling strength $J_z/t$ is also investigated as captured in Fig. 4. Here, the weak infinitesimal growth of the energy of the hole with the system size is a general characteristic of all plots of $E_h/t$ versus N, for fixed values of $J_z/t$. However, The dependence of $E_h/t$ on N for small N in the strong coupling regime is stronger than its dependence in the weak coupling regime. This is because for relatively small distances (small N), the string connecting the holon to the spinon in the strong coupling regime is stronger than the weak coupling regime. At large N, corresponding to large distances, $E_h/t$ is found to assume a constant value, indicating the complete separation of holon from spinon.

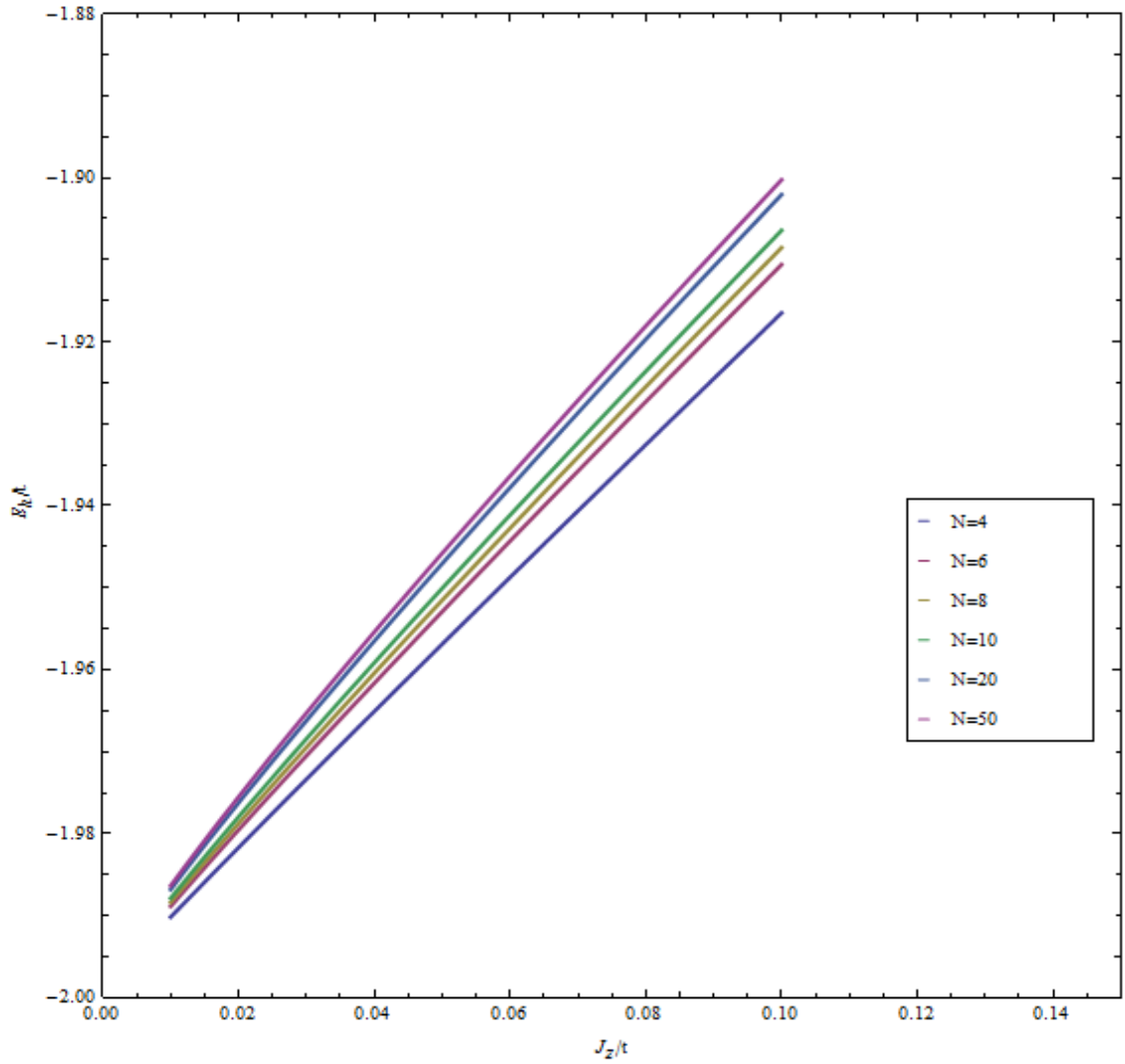

Fig.3. The energy of the hole in the weak Coupling regime in 1D for N=4 to 50. The energy of the hole increases infinitesimally from N=4 to 50. The increase in the energy of the hole between N=20 and N=50 appears to be insignificant, thus implying that the hole is on the verge of escaping from the confining potential.

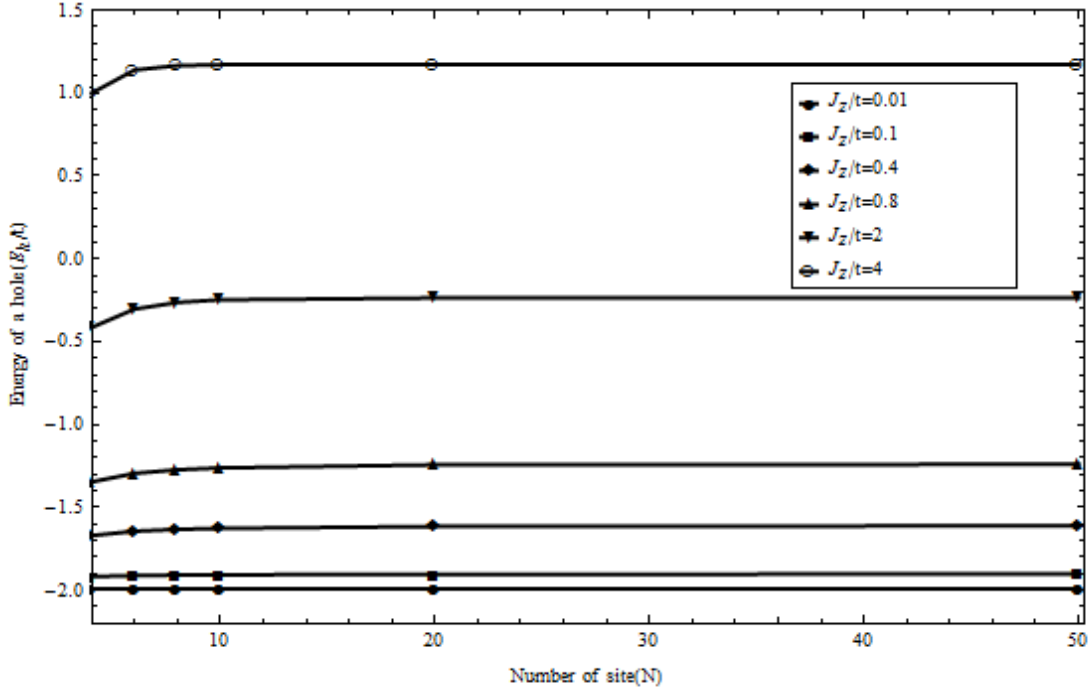

Fig. 4. The energy of a hole as a function of the number of sites N for fixed values of the coupling strength. The weak dependence of $E_h/t$ on $J_z/t$ for small N is found to vanish at large N.

**4.3 Strong coupling regime**

In this regime, the energy of the hole can approximately be described by the equation:

$$E_h = J_z - \frac{39t^2}{10J_z} \qquad (22)$$

Here, the dependence of $E_h$ on $J_z$ in this regime is discussed. For comparison, both the result obtained from the exact diagonalization of finite systems in the strong coupline regime, as shown in Table 1 and that in eqn. (22) are depicted in Fig. 5. As observed in Fig. 5, agreement between these two presentations is excellent, especially in the limit $J_z/t \gg 1$, where they both converged. In this regime, eqn. (22) predicts that the energy of the hole is independent of the system size. Since an increase in $J_z$ promotes strong magnetic coupling between neighbouring spins, the motion of the hole becomes strongly incoherent in this regime. Accordingly, the velocities corresponding to charge and spin degrees of freedom become comparable. In other words, the spinon appears to be strongly couple to the holon. Under this situation, the hole may be treated as a localized particle. In two dimensional antiferromagnet, the number of "anti-Neel spins", which is a measure of the size of the polaron decreases significantly.

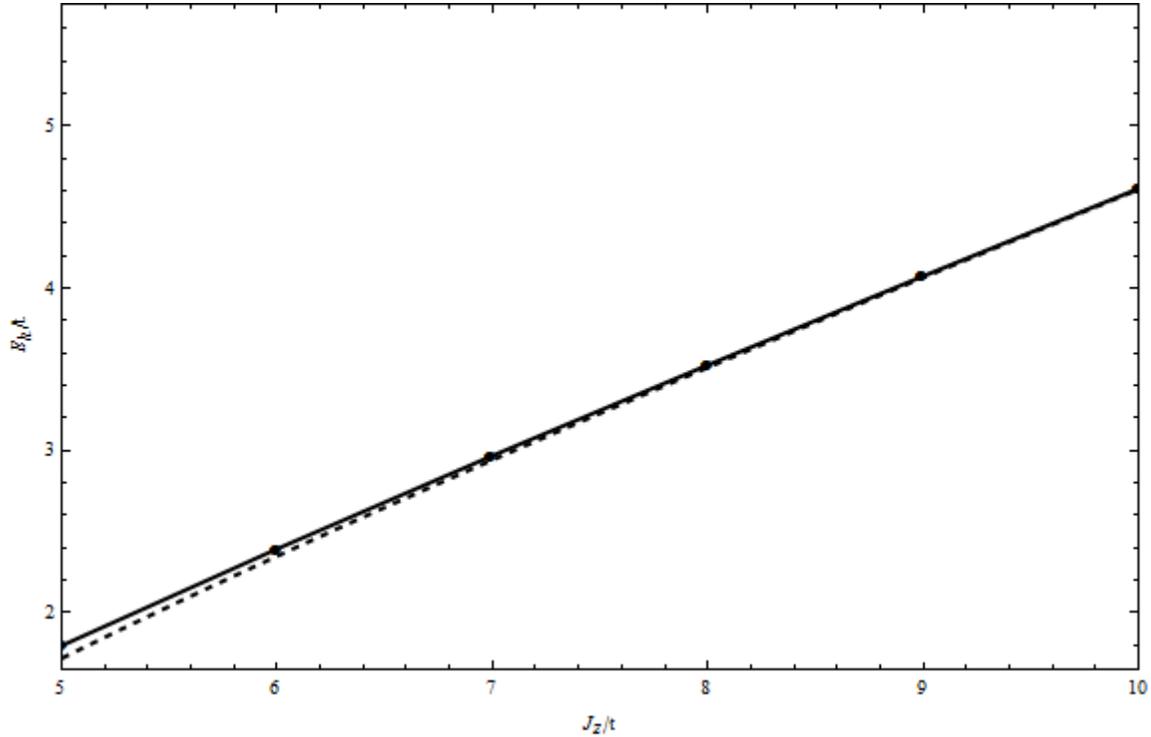

Fig. 5. The energy of a hole as a function of the coupling strength. The dotted line represents eqn.(22), while the thick line is obtained from the exact diagonalization of finite clusters.

### 4.4. Experimental realization of spin-charge separation in 1d cuprates

Experimental evidence for spin-charge separation is provided by ARPES study of the 1D cuprate $Sr_2CuO_3$ as shown in Figs.6 and 7 [8]. The ARPES spectrum at the $\Gamma$ point (K=0) has a very distinct peak at 1.2eV below the Femi energy $E_F$, while a rapidly dispersing small structure appears around 1.5–2.0eV (see Fig.6). These two structures merge to form a prominent peak at 0.8 eV at $kb/\pi =0.5$. The rapidly dispersing branch is symmetric about $kF$ (see Fig.7) and arises from holon states, whereas the peak at lower binding energies is due to spinons and is observable only in the first half of the Brillouin zone. The disappearance of the spinon structure in the second half of the Brillouin zone and the presence of holon in this zone indicate complete separation of hole from spin.

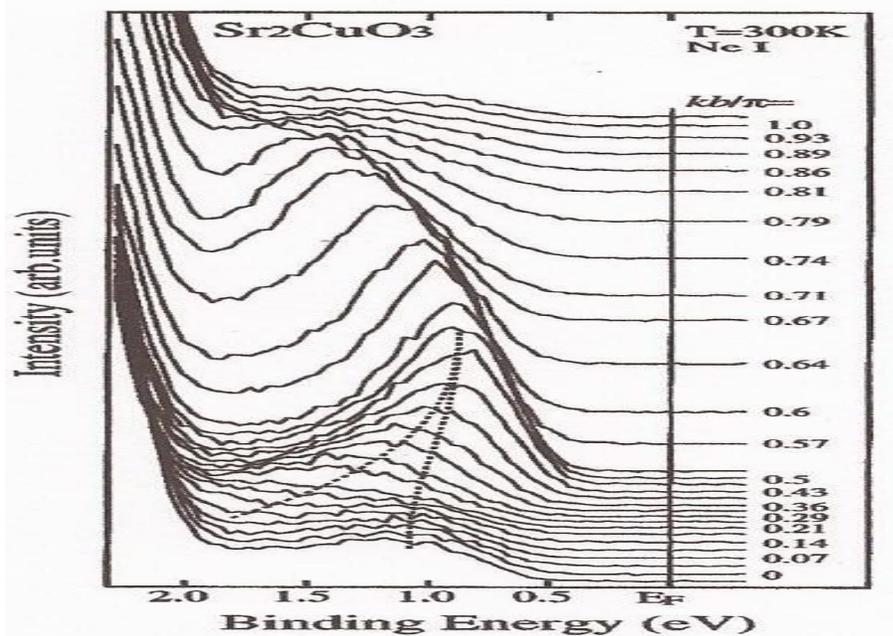

Fig.6. ARPES spectra of $Sr_2CuO_3$ near EF measured along the b chain direction using Ne I photons. The distinct peak represents the holon, while the rapidly dispersing structure represents the spinon. Source: Ref. [8].

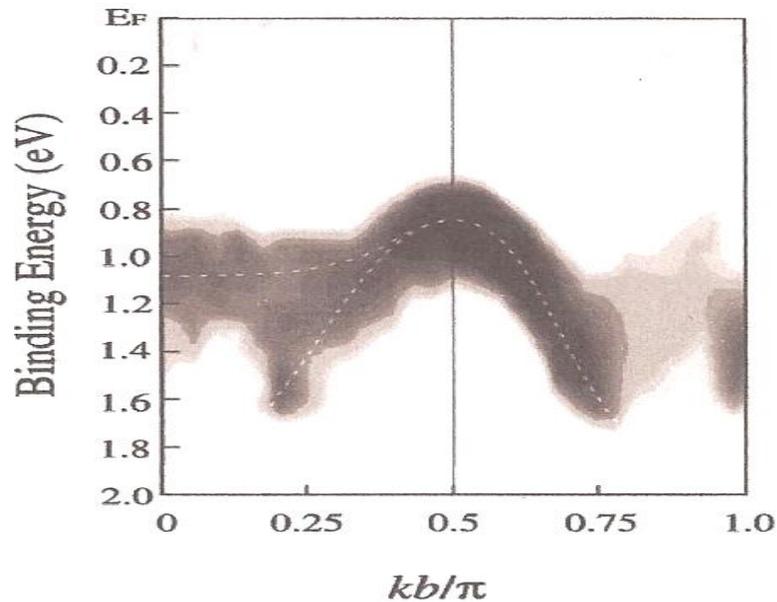

Fig.7. Band dispersion of $Sr_2CuO_3$ obtained from ARPES measurement. The weakly propagating structure representing the spinon is evident in the first half of the Brillouin zone and quickly disappears in the second half of the Brillouin zone. The strongly propagating structure representing the holon structure survives the second part of the Brillouin zone. Source: Ref. [8].

# 6. Conclusions

Exact diagonalization study on doped 1D cuprates showed that the hole is not completely decoupled from the ferromagnetic bond (spinon), but experiences some amount of linear "string" potential that tends to compromise its motion. This is in contrast to what is obtained in the bulk limit. However, for large N (corresponding to bulk limit), the hole is found to detach from this string and propagates as a free particle. Hence, two independent quasipaticles, namely spin (spinons) and charge (holons) are produced. The appearance of these two independent quasi-particles provides further evidence for spin-charge separation in the bulk limit in agreement with both theoretical predictions and experimental results [8-10, 12, 29]. In the strong coupling regime, the energy of the hole is found to be independent of the system size. An increase in $J_z$ increases the magnetic energy cost. In other words, the quasiparticles, namely spinon and holon appear to be strongly coupled to each other. Under this situation, the hole may be treated as a localized particle.

The problem of spin-charge separation in 1D quantum fluids was recently investigated in ref. [30]. Within the luttinger liquid theory, the authors formulated the dynamical response in terms of noninteracting bosonic collective excitations carrying either charge or spin and argued that as a result of spectral nonlinearity, long-lived excitations are best understood in terms of generally strongly interacting fermionic holons and spinons. The concept of spincharge separation mechanism has also been introduced to the quark–lepton unification models which consider the lepton number as the fourth color. In certain finite-density systems, quarks and leptons are found to decompose into spinons and chargons, which carry the spin and charge degrees of freedom respectively [31]. The interest in the study of luttinger liquid systems is motivated by potential applications in spintronics and quantum computation. For example, a striking example of Luttinger liquid being currently investigated because of it possible application in spintronics is the helical Luttinger liquid. It is characterized by spin momentum locking (i.e. particles with spin up and down with respect to a well-defined *z*-axis propagate in opposite directions) [32].